\documentclass{elsart}

\usepackage{graphicx}

\def\'{'}

\begin{document}

\begin{frontmatter}

%\draft
\title{Stretched-Gaussian asymptotics of the truncated 
L\'evy flights for the diffusion in nonhomogeneous media}

\author
{Tomasz Srokowski}

\address{
 Institute of Nuclear Physics, Polish Academy of Sciences, PL -- 31-342
Krak\'ow, Poland }

\begin{abstract}

The L\'evy, jumping process, defined in terms of the jumping size 
distribution and the waiting time distribution, is considered. The 
jumping rate depends on the process value. The fractional diffusion 
equation, which contains the variable diffusion coefficient, 
is solved in the diffusion limit. That solution resolves itself 
to the stretched Gaussian when the order parameter $\mu\to2$. 
The truncation of the L\'evy flights, in the exponential and power-law 
form, is introduced and the corresponding random walk process 
is simulated by the Monte Carlo method. The stretched Gaussian tails 
are found in both cases. The time which is needed to reach the limiting 
distribution strongly depends on the jumping rate parameter. When the 
cutoff function falls slowly, the tail of the distribution appears 
to be algebraic. 

\end{abstract}

\begin{keyword}
Diffusion; Fractional equation; Truncated L\'evy flights

PACS 02.50.Ey, 05.40.Fb, 05.60.-k
\end{keyword}

\end{frontmatter}

\maketitle

\section{Introduction}

Transport in physical systems can be described in terms of a jumping process 
which is completely defined by two probability distributions. They determine 
spatial and temporal characteristics of the system and are called the 
jumping size, $\lambda(x)$, and waiting time, $w(t)$, probability distributions, 
respectively. The stochastic trajectory consists of infinitely fast jumps 
which are separated by intervals when 
the Brownian particle is at rest (e.g., due to traps). Usually, 
the distributions $\lambda$ and $w$ are regarded as mutually independent 
and processes which they generate are studied in the framework of the decoupled 
continuous time random walk (CTRW). If the distribution $w(t)$ is Poissonian, 
the jumps take place at random, uniformly distributed times and the 
corresponding process is Markovian. The algebraic form of $w(t)$, which 
possesses long tails, is of particular interest; it leads to long rests and 
then to a weak, subdiffusive transport \cite{met}. The jumping size 
distribution $\lambda$ can assume -- in order to be stable -- either 
the Gaussian form, or obey the general L\'evy distribution 
which corresponds to processes with long tails, frequently encountered in many 
areas of physics, sociology, medicine \cite{wes} and many others. In the 
diffusion limit of small 
wave numbers, the master equation for the decoupled CTRW, 
and for the case of the L\'evy distributed jumping size, resolves itself to 
the fractional diffusion equation with a constant diffusion coefficient.

However, many physical phenomena require taking into account that the space has 
some structure and that the diffusion coefficient must actually be variable. 
It is the case when one considers the transport in porous, inhomogeneous media 
and in plasmas, as well as for the L\'evy flights in systems with an external 
potential \cite{bro}. In the Hamiltonian systems, 
the speed of transport depends on regular structures in the phase space 
\cite{zas}. Any description of the diffusion on fractals must 
involve the variable diffusion coefficient, and that one in the power-law 
form \cite{osh,met4}. Similarly, the diffusion on multifractal structures can be 
regarded as a superposition of the solutions which correspond to the 
individual fractals \cite{sro1}. The power-law form of the diffusion 
coefficient has been also used to describe
e.g. the transport of fast electrons in a hot plasma \cite{ved}
and the turbulent two-particle diffusion \cite{fuj}. 
The presence of long jumps indicates a high complexity and the 
existence of long-range correlations. One can expect that especially in 
such systems the waiting time depends on the position. The jumping process, 
stationary and Markovian, which takes into account 
that dependence, has been proposed in Ref.\cite{kam}: the distribution 
$w(t)$ is Poissonian with a $x$-dependent jumping rate $\nu(x)$. 
In this paper, we consider that process for the L\'evy distributed 
$\lambda(x)$ and solve the corresponding fractional equation. 
The solution consistently takes into account that the symmetric 
L\'evy distribution, defined by the characteristic 
function $\exp(-|k|^\mu)$, $0<\mu\le2$, exhibits two qualitatively different 
kinds of behaviour in its asymptotic limit: it is either 
algebraic, $\sim |x|^{-\mu-1}$, for $0<\mu<2$ or Gaussian for $\mu=2$. 

The L\'evy process is characterized by very long jumps since the tails 
of the L\'evy distribution fall slowly and the 
second moment is divergent. In a concrete, realistic problem, however, 
the available space is finite and the asymptotics differs from the L\'evy 
tail. One can take that into account by introducing a cutoff 
to the stochastic equations by multiplying the jumping size 
distribution -- in the L\'evy form -- by a function which falls not slower 
than $1/|x|^3$, usually by the exponential or the algebraic function. Since 
the resulting probability distribution has the finite variance, in the 
diffusion approximation the process is equivalent to the Gaussian limit 
$\mu\to2$. 

Restrictions on the jump size are necessary also for problems connected 
with the random transport of massive particles. Since the velocity of 
propagation is then finite, the jump, which takes place within a given 
time interval, must have a finite length. If the jump size is 
governed by the particle velocity, a coupling between spatial and temporal 
characteristics of the system emerges in the framework of CTRW. That coupled 
form of the CTRW is known as the L\'evy walk \cite{gei,zum,kla1}. It has many 
applications, e.g. for the turbulence \cite{shl} and the chaotic diffusion 
in Josephson junctions \cite{gei}. 

The aim of this paper is to study the limit in which the algebraic 
tail of the L\'evy process becomes the exponential function. 
Utilizing the result for $\mu\to2$ can improve the 
solution of the fractional equation itself and it is suited to describe 
the truncated L\'evy flights in the diffusion approximation. 
The paper is organized as follows.
In Sec.II we solve the fractional equation and derive the stretched Gaussian 
asymptotics in the limit $\mu\to2$. Sec.III is devoted to the truncated 
L\'evy flights: we present the probability distributions, obtained from 
simulations of random walk trajectories, for both exponential and power-law 
form of the cutoff. The results are summarized in Sec.IV.

\section{The stretched-Gaussian limit of the L\'evy process}

We consider the random walk process defined by the waiting time probability 
distribution $w(t)$ and the jump-size distribution $\lambda(x)$. They 
are of the form
\begin{equation}
\label{poi}
w(t)=\nu(x){\mbox e}^{-\nu(x)t},
\end{equation}
where $\nu(x)=|x|^{-\theta}$ ($\theta>-1$) and 
\begin{equation}
\label{lev}
\lambda(x)=\sqrt{2/\pi} \int_0^\infty \exp(-K^\mu k^\mu)\cos(kx)dk,
\end{equation}
respectively. The latter expression corresponds to the symmetric L\'evy 
distribution. The master equation for the above process is of the form 
\cite{kam}
  \begin{equation}
  \label{fpkp}
  \frac{\partial}{\partial t}p(x,t) = -\nu(x)p(x,t) +
  \int \nu(x') \lambda(|x-x'|) p(x',t) dx'.
  \end{equation}
In the diffusion limit of small wave numbers, 
the Eq.(\ref{fpkp}) can be reduced to 
the fractional equation. Indeed, the Fourier transform of jump-size 
distribution, 
${\widetilde \lambda}(k)=\exp(-K^\mu |k|^\mu)$, can be expanded ${\widetilde 
\lambda}(k)= 1-K^\mu |k|^\mu+\dots$ and the master equation in the Fourier space 
becomes the following equation \cite{sro}
\begin{equation}
\label{fracek}
\frac{\partial{\widetilde p}(k,t)}{\partial t}=-K^\mu |k|^\mu{\widetilde F}
[\nu(x)p(x,t)].
\end{equation}
The above approximation means that the summation over jumps is substituted by 
the integral and it agrees with the exact result if the jumps are 
small, compared to the entire trajectory. One can demonstrate by 
estimating the neglected terms in the Euler-Mclaurin summation formula 
that the approximation fails near the origin and for $\mu\le1$ \cite{bark}. 
In the other cases, the solution of the fractional equation converges with 
time to the exact result. Therefore, in the following, we assume $\mu>1$. 

The inversion of the Fourier transforms in Eq.(\ref{fracek}) yields 
the fractional equation
\begin{equation}
\label{fraced}
\frac{\partial p(x,t)}{\partial t}=
K^\mu\frac{\partial^\mu[\nu(x)p(x,t)]}{\partial|x|^\mu}, 
\end{equation}
where $\partial^\mu/\partial|x|^\mu$ is the Riesz-Weyl derivative, defined 
for $1<\mu<2$ in the following way \cite{che2}
\begin{equation}
\label{fracdef}
\frac{\partial^\mu}{\partial|x|^\mu}f(x,t)=\frac{-1}{2\cos(\pi\mu/2)
\Gamma(2-\mu)}\frac{\partial^2}{\partial x^2}\int_{-\infty}^\infty
\frac{f(x',t)}{|x-x'|^{\mu-1}}dx'.
\end{equation}

Eq.(\ref{fraced}) for the constant diffusion coefficient --  
often generalized to take into account non-Markovian features of the 
trapping mechanism in the framework of CTRW by substituting the 
simple time derivative by the fractional Riemann-Liouville 
derivative \cite{met} -- is frequently considered and solved by means of a 
variety of methods. The solution resolves itself to the L\'evy 
stable distribution with the asymptotic power-law $x$-dependence 
and divergent second moment. The method of solution which is especially 
interesting for our considerations involves the Fox functions. 
The well-known result of Schneider \cite{sch} states that any L\'evy 
distribution, both symmetric and asymmetric, can be expressed as 
$H_{2,2}^{1,1}$. In order to solve the Eq.(\ref{fraced}) for the case 
of the variable jumping rate, $\nu(x)=|x|^{-\theta}$, we conjecture 
that the solution also belongs 
to the class of functions $H_{2,2}^{1,1}$ and it has the scaling 
property. Then the ansatz is the following
\begin{eqnarray} 
\label{s1}
p(x,t)=Na(t) H_{2,2}^{1,1}\left[a(t) |x|\left|\begin{array}{c}
(a_1,A_1),(a_2,A_2)\\
\\
(b_1,B_1),(b_2,B_2)
\end{array}\right.\right],
\end{eqnarray}
where $N$ is the normalization constant. The Fox function is 
defined in the following way \cite{fox,mat}
  \begin{eqnarray} 
\label{a1}
H_{pq}^{mn}\left[z\left|\begin{array}{c}
(a_p,A_p)\\
\\
(b_q,B_q)
\end{array}\right.\right]=\frac{1}{2\pi i}\int_L\chi(s)z^sds,
  \end{eqnarray}
where 
\begin{equation}
\label{a2}
\chi(s)=\frac{\prod_1^m\Gamma(b_j-B_js)\prod_1^n\Gamma(1-a_j+A_js)}
{\prod_{m+1}^q\Gamma(1-b_j+B_js)\prod_{n+1}^p\Gamma(a_j-A_js)}.
\end{equation}
The contour $L$ separates the poles belonging to the 
two groups of the gamma function in the Eq.(\ref{a2}). 
Evaluation of the residues 
leads to the well-known series expansion of the Fox function:
  \begin{eqnarray}  
\label{a3}
\hskip-1cm 
H_{pq}^{mn}\left[z\left|\begin{array}{c}
(a_p,A_p)\\
\\
(b_q,B_q)
\end{array}\right.\right]=\sum_{h=1}^m\sum_{\nu=0}^\infty\frac{\prod_{j=1,j\ne
h}^m\Gamma(b_j-B_j\frac{b_h+\nu}{B_h})\prod_{j=1}^n\Gamma(1-a_j+A_j\frac{b_h+\nu
}{B_h})}{\prod_{j=m+1}^q\Gamma(1-b_j+B_j\frac{b_h+\nu}{B_h})\prod_{j=n+1}^p
\Gamma(a_j-A_j\frac{b_h+\nu}{B_h})}\frac{(-1)^\nu z^{(b_h+\nu)/B_h}}{\nu!B_h},
  \end{eqnarray}

We will try to solve the fractional equation (\ref{fraced}) 
by inserting the function (\ref{s1}). This procedure, if successful, 
would allow us to find conditions for the coefficients and the 
function $a(t)$. The idea to assume the solution in the form 
(\ref{s1}) is motivated by the following property of the 
Fox function
  \begin{eqnarray} 
\label{a5}
z^\sigma H_{pq}^{mn}\left[z\left|\begin{array}{c}
(a_p,A_p)\\
\\
(b_q,B_q)
\end{array}\right.\right]= 
H_{pq}^{mn}\left[z\left|\begin{array}{c}
(a_p+\sigma A_p,A_p)\\
\\
(b_q+\sigma B_q,B_q)
\end{array}\right.\right]
  \end{eqnarray}
which eliminates the algebraic factor by means of a 
simple shift of the coefficients. One can demonstrate that 
Eq.(\ref{fraced}) cannot be satisfied, in general, by the function 
(\ref{s1}) for any choice of the parameters. However, as long as 
we are interested only in the diffusion limit of small wave numbers 
$|k|$ (large $|x|$), the higher terms in the characteristic function 
expansion can be neglected. Therefore we require that the 
Eq.(\ref{fraced}) should be satisfied by a function which agrees 
with the exact solution only up the terms of the order $|k|^\mu$ 
in the Fourier space. Note that this condition 
does not introduce any additional idealization since the 
Eq.(\ref{fraced}) itself has been constructed on the same assumption: 
the higher terms in the $|k|$ expansion of $\lambda(x)$ 
have been also neglected. 

First, we need the Fourier transform of the Fox function. Since 
the process is symmetric, we can utilize the formula for the 
cosine transform which yields also a Fox function but of the 
enhanced order: 
  \begin{eqnarray} 
\label{a6}
\hskip-1cm
\int_0^\infty H_{pq}^{mn}\left[x\left|\begin{array}{c}
(a_p,A_p)\\
\\
(b_q,B_q)
\end{array}\right.\right]\cos(kx)dx= 
\frac{\pi}{k}H_{q+1,p+2}^{n+1,m}\left[k\left|\begin{array}{l}
(1-b_q,B_q),(1,1/2)\\
\\
(1,1),(1-a_p,A_p),(1,1/2)
\end{array}\right.\right].
  \end{eqnarray}
The function ${\widetilde p}(k,t)$ is then of the order 
$H_{3,4}^{2,1}$. The Fourier transform of the function 
$p_\theta=x^{-\theta} p(x,t)$ can be obtained in 
the same way. Next, we 
insert the appropriate functions into the Eq.(\ref{fracek}) 
and expand both sides of the equation by using the formula (\ref{a3}). 
Let us denote the expansion coefficients of the functions 
${\widetilde p}(k,t)$ and ${\widetilde p_\theta}(k,t)$ by $h_{\sigma,\nu}$ 
and $h_{\sigma,\nu}^{(\theta)}$, respectively; $\sigma$ assumes 
the values 1 and 2. Applying the Eq.(\ref{a3}) yields
\begin{equation}
\label{fl}
{\widetilde p}(k,t)=h_{1,0}+h_{1,1}|k|+h_{2,0}|k|^{(1-a_1)/A_1-1}+h_{2,1}
|k|^{(2-a_1)/A_1-1}+h_{1,2}k^2+\dots
\end{equation}
and
\begin{equation}
\label{fp}
{\widetilde p_\theta}(k,t)=h_{1,0}^{(\theta)}+h_{1,1}^{(\theta)}|k|+
h_{2,0}^{(\theta)}|k|^{(1-a_1+\theta A_1)/A_1-1}+\dots.
\end{equation}
After inserting the above expressions to the Eq.(\ref{fracek}), 
we find some simple relations among the coefficients of the Fox function 
by comparison of the exponents. To get the term $|k|^\mu$ on lhs, which 
corresponds to the term $k^0$ on rhs, we need the condition 
$(2-a_1)/A_1-1=\mu$. Moreover, we attach the third term on rhs to the first 
one by putting $(1-a_1+\theta A_1)/A_1-1=0$; it is not possible to balance 
the third term by another one on the lhs. The above conditions determine 
two coefficients of the Fox function: $a_1=1-(1-\theta)/(\mu+\theta)$ 
and $A_1=1/(\mu+\theta)$.
The coefficients $h_{1,1}$ and $h_{1,1}^{(\theta)}$ vanish identically 
since the gamma function in the denominators has its argument equal 0. The 
only remaining term can be eliminated by assuming the condition 
$1-b_2-B_2(1-\theta)=0$; then $h_{2,0}=0$ since that term contains the function 
$\Gamma(1-b_2-B_2(1-\theta))$ in the denominator. Therefore, such choice of 
the coefficients reduces the function (\ref{fl}) to the two terms and it makes 
it 
identical with the probability distribution for the L\'evy process in the 
diffusion limit. Finally, the 
Eq.(\ref{fracek}) becomes a simple differential equation which determines 
the function $a(t)$:
\begin{equation}
\label{fract}
\dot\xi=K^\mu\frac{h_0^{(\theta)}}{h_{2,1}}\xi^{-\theta/\mu},
\end{equation}
where $h_0^{(\theta)}=h_{1,0}^{(\theta)}+h_{2,0}^{(\theta)}$ and 
$\xi(t)=a^{-\mu}$. The solution 
\begin{equation}
\label{solt}
a(t)=\left[K^\mu\frac{h_0^{(\theta)}}{h_{2,1}}
\left(1+\frac{\theta}{\mu}\right)t\right]^{-1/(\mu+\theta)}
\end{equation}
corresponds to the initial condition $p(x,0)=\delta(x)$. 
The coefficient $h_{2,1}$ can be determined directly 
from Eq.(\ref{a3}), whereas 
$h_0^{(\theta)}=(2\pi)^{-1}a^{-\theta}{\widetilde p}(0,t)=\pi^{-1}a^{-\theta}
\int_0^\infty p(x,t)dx$ can be expressed in terms of the Mellin transform 
from the Fox function, $\chi(s)$, and then easily evaluated. 

In the limit $\mu\to2$, the asymptotic behaviour of the fractional 
equation changes qualitatively. It is no longer algebraic; the tails 
of the distribution, and consequently the tails of the Fox function, 
have to assume the exponential form for $\mu=2$. It is possible only 
if the algebraic contribution from the residues vanishes. If $n=0$, 
all poles are outside the contour $L$ and $H_{p,q}^{m,0}$ is 
given by the integral over a vertical straight line:
  \begin{eqnarray} 
\label{bmi}
H_{p,q}^{m,0}=\frac{1}{2\pi i}\int_{w-i\infty}^{w+i\infty}\chi(s)z^sds, 
  \end{eqnarray}
where $w<Re(b_i/B_i)$. The above integral is usually neglected if $n>0$ 
because it is small compared to the contribution from the residues. 

The Fox function in the required form can be obtained from (\ref{s1}) 
by applying the reduction formula if 
the coefficients in the main diagonal are 
equal. This demand imposes an additional condition on $b_2$ and $B_2$ 
which allows us to determine these coefficients: 
$b_2=1-(1-\theta)/(2+\theta)$ and $B_2=1/(2+\theta)$.
Note that the above choice of $(b_2,B_2)$ yields 
$h_{1,2}=0$ in Eq.(\ref{fl}) and then the solution 
of Eq.(\ref{fraced}) is exact up to the order $k^2$ for any $\mu$.
The most general solution of the fractional equation in the form of 
the function $H^{1,1}_{2,2}$, which involves the 
required conditions, is the following 
\begin{eqnarray} 
\label{smu}
p(x,t)=Na(t) H_{2,2}^{1,1}\left[a(t) |x|\left|\begin{array}{c}
(1-\frac{1-\theta}{\mu+\theta},\frac{1}{\mu+\theta}),(a_2,A_2)\\
\\
(b_1,B_1),(1-\frac{1-\theta}{2+\theta},\frac{1}{2+\theta})
\end{array}\right.\right].
\end{eqnarray}
The coefficients in the main diagonal have a simple interpretation. 
Applying Eq.(\ref{a3}) to (\ref{smu}) reveals that $p(x,t)$ 
behaves as $|x|^{b_1/B_1}$ for $|x|\to 0$ and then the parameters 
$b_1$ and $B_1$ are responsible for the shape of the probability 
distribution near the origin. On the other hand, the parameters $a_1$ 
and $A_1$ determine the asymptotic shape of the distribution. It can 
be demonstrated by applying the following property of the Fox function
  \begin{eqnarray} 
\label{a4}
H_{pq}^{mn}\left[z\left|\begin{array}{c}
(a_p,A_p)\\
\\
(b_q,B_q)
\end{array}\right.\right]= 
H_{pq}^{mn}\left[\frac{1}{z}\left|\begin{array}{c}
(1-b_q,B_q)\\
\\
(1-a_p,A_p)
\end{array}\right.\right]
  \end{eqnarray}
and by expansion according to Eq.(\ref{a3}): the leading term is of the 
form $p(x,t)\sim |x|^{(2-a_1)/A_1}=|x|^{-1-\mu}~~~~(|x|\to\infty)$. 
Now it becomes clear why our method of solving the Eq.(\ref{fraced}) 
did not determine the parameters $(b_1,B_1)$. Since we neglected 
higher terms in the $k$-expansion, the region of small $|x|$ remained 
beyond the scope of the approximation. However, we can supplement that 
solution by referring directly to the master equation 
(\ref{fpkp}) which reveals a simple 
behaviour near the origin. That equation is satisfied by the stationary 
solution $1/\nu(x)=|x|^\theta$, for any normalizable $\lambda(x)$. 
Obviously, such $p(x,t)$ cannot be interpreted as the probability density 
distribution since the normalization integral diverges in infinity 
but it properly reproduces that distribution 
for small $|x|$. Therefore we obtain the additional condition 
$b_1=\theta B_1$ which improves the agreement of our solution with the 
solution of the master equation.
The probability distribution for the case $\theta=0$ can be 
easily found by the direct solution of the fractional equation which is 
exact and yields the following values of the parameters: $b_1=0$, $B_1=1$, 
$a_2=1/2$, and $A_2=1/2$.

In the case $\mu=2$, the main diagonal in Eq.(\ref{smu}) can be 
eliminated and the solution of the fractional 
equation (\ref{fraced}) in the limit $\mu\to 2$ takes the form
\begin{eqnarray} 
\label{s1r}
p(x,t)=Na(t) H_{1,1}^{1,0}\left[a(t) |x|\left|\begin{array}{c}
(a_2,A_2)\\
\\
(b_1,B_1)
\end{array}\right.\right].
\end{eqnarray}
The function $a(t)$ is given by Eq.(\ref{solt}) in the following form
\begin{equation}
\label{solte}
a(t)=\left[K^2(2+\theta)\frac{h_0}{h_2}t\right]^{-1/(2+\theta)},
\end{equation}
where $h_0=\Gamma(b_1+B_1(1-\theta))/\Gamma(a_2+A_2(1-\theta))$ and 
$h_2=\Gamma(b_1+3B_1)/\Gamma(a_2+3A_2)$ are the coefficients of the 
$k$-expansion of the functions $|x|^{-\theta}p(x,t)$ and $p(x,t)$, 
respectively. The asymptotic expression for $p(x,t)$ can be obtained 
from the estimation of the integral (\ref{bmi}) by the method of 
steepest descents \cite{bra}. The result reads
  \begin{equation} 
\label{solex}
H_{1,1}^{1,0}(z)\approx{\mbox e}^{i\pi(\alpha'-1/2)}E(z{\mbox e}^{i\pi\alpha}),
  \end{equation}
where
  \begin{equation} 
\label{solex1}
E(z)=\frac{1}{2\pi i\alpha}\sum_{j=0}^\infty C_j(\beta\alpha^\alpha z)^
{(1-\alpha'-j)/\alpha}\exp(\beta\alpha^\alpha z)^{1/\alpha}
  \end{equation}
and $\alpha=B_1-A_2$, $\alpha'=a_2-b_1+1/2$, $\beta=A_2^{A_2}/B_1^{B_1}$, 
$z=a|x|$. 
The final result appears to be a stretched Gaussian, modified by an 
algebraic factor and a series which converges to a constant for 
$z\to\infty$:
  \begin{equation} 
\label{solex2}
H_{1,1}^{1,0}(z)\approx\frac{1}{2\pi\alpha}z^{(1-\alpha')/\alpha}
\exp(-\beta^{1/\alpha}\alpha z^{1/\alpha})\beta^{(1-\alpha')/\alpha}
\alpha^{1-\alpha'}\sum_{j=0}^\infty (-1)^j C_j\beta^{-j/\alpha}\alpha^{-j} 
z^{-j/\alpha}.
  \end{equation}
The above expression has been obtained by Wyss \cite{wys} as the expansion 
of $H^{3,0}_{2,3}(z)$ which satisfies a generalized diffusion equation. 
It is the integral equation in respect to the time 
variable (non-Markovian) which resolves itself to the 
fractional diffusion equation; that equation is commonly used 
to handle the subdiffusive 
processes in the framework of the CTRW \cite{zum,kla,met}. 
The function which contains the stretched Gaussian, modified by 
the power-law factor, is used as the asymptotic form of the propagator 
for diffusion on fractals, e.g. on the Sierpi\'nski gasket \cite{rom}. 

The coefficients $C_j$ are defined by means of the following expression
\begin{equation}
\label{cj}
\frac{\Gamma(1-a_2+A_2s)}{\Gamma(1-b_1+B_1s)}(\beta\alpha^\alpha)^{-s}
\equiv G=\sum_{j=0}^\infty\frac{C_j}{\Gamma(\alpha s+\alpha'+j)}.
\end{equation}
They can be explicitly evaluated by a subtraction of the consecutive 
terms and by taking the limit $s\to\infty$. More precisely, $C_j$ are 
given by the following recurrence formula
\begin{eqnarray}
\label{crec}
\hskip-1cm
C_j&=&\lim_{s\to\infty}\left[G\Gamma(\alpha s+\alpha')-C_0-\frac{C_1}{\alpha 
s+\alpha'}
-\dots-\frac{C_{j}}{(\alpha s+\alpha')
%(\alpha s+\alpha'+1)
\dots(\alpha s+\alpha'+j-1)}\right]\times\nonumber\\
&\nonumber\\
&\times&(\alpha s+\alpha')(\alpha s+\alpha'+1)
\dots(\alpha s+\alpha'+j-1)
\end{eqnarray}
where 
\begin{equation}
\label{c0}
C_0=\sqrt{2\pi}\alpha^{\alpha'-1/2}A_2^{1/2-a_2}B_1^{b_1-1/2}
\end{equation}
has been obtained from the expansion of the gamma functions 
by means of the Stirling formula. The exponent of the stretched Gaussian, 
$1/\alpha$, is connected 
with higher moments of the distribution $p(x,t)$ and it cannot be 
determined in the framework of the diffusion approximation. 

All moments of the distribution $p(x,t)$ are convergent. In particular,
the variance, which determines the diffusion properties of the system, 
is given by the expansion coefficient $h_2$ in a simple way:
\begin{equation}
\label{var}
\langle x^2\rangle=-\frac{\partial^2}{\partial k^2}{\widetilde p}(0,t)=
h_2a^{-2}\sim t^{\frac{2}{2+\theta}}. 
\end{equation}
For $\theta=0$ the diffusion coefficient 
$D=\lim_{t\to\infty}\langle x^2\rangle(t)/2t$ assumes a finite value 
and the diffusion is normal. The case $\theta\ne 0$ 
means the anomalous diffusion: 
either the enhanced one for $\theta<0$, or the subdiffusion for $\theta>0$; 
the diffusion coefficients are then $\infty$ or 0, respectively. The kind of 
diffusion depends only on $\theta$ and it is not sensitive on free parameters. 
The anomalous diffusion is frequently encountered in physical phenomena, in 
particular in complex and disordered systems \cite{bou}, as well as in 
dynamical systems \cite{zas}. 

On the other hand, the fractional equation (\ref{fraced}) 
for $\mu=2$ assumes a form of the simple diffusion equation 
\begin{equation}
\label{fraceg}
\frac{\partial p(x,t)}{\partial t}=
K^2\frac{\partial^2[|x|^{-\theta}p(x,t)]}{\partial x^2} 
\end{equation}
which can be solved 
exactly just by assuming the scaling form of the solution 
$p(x,t)=a(t)f(a(t)x)$. The functions $a(t)$ 
and $f(ax)$ are derived by inserting to the Eq.(\ref{fraceg}) and by 
separation of the variables \cite{kwo}. That procedure finally yields
   \begin{equation}
   \label{kwok}
   p(x,t)=N(K^2t)^{-\frac{1+\theta}{2+\theta}}|x|^{\theta}
\exp\left(-\frac{|x|^{2+\theta}}{K^2 (2+\theta)^2t}\right).
   \end{equation}
Eq.(\ref{fraceg}) follows from the 
master equation (\ref{fpkp}) with the Gaussian $\lambda(x)$, when 
one neglects all terms higher than the second one in the 
Kramers-Moyal expansion. That procedure is justified if jumps are small 
and $\nu(x)$ is a smooth function \cite{vkam}. 
We can expect that the solution of 
the master equation converges with time to Eq.(\ref{kwok}) 
and this convergence is fast for small $|\theta|$. 
Convergence of the tails must be slow, because the contribution from 
large jumps is substantial for large $|x|$. 
Indeed, we will demonstrate in the following that for large $|\theta|$, 
in particular for $\theta$ close to $-1$, a very long time is required. 

The result (\ref{kwok}) is useful for further improvement of the solution 
(\ref{smu}). The comparison of Eqs. (\ref{solex2}) and (\ref{kwok}) 
yields the conditions for the Fox function coefficients which ensure 
the proper limit $\mu\to2$: $\lim_{\mu\to2}(B_1-A_2)=1/(2+\theta)$ and 
$\lim_{\mu\to2}(1-\alpha')/\alpha=\theta$. By inserting the coefficients 
which follow from those limiting values to the Eq.(\ref{smu}) and by 
assuming, in addition, that $B_1=1$, we obtain a particular solution 
of the fractional equation (\ref{fraced}) in the form
\begin{eqnarray} 
\label{smu1}
p(x,t)=Na(t) H_{2,2}^{1,1}\left[a(t) |x|\left|\begin{array}{c}
(1-\frac{1-
\theta}{\mu+\theta},\frac{1}{\mu+\theta}),(\frac{1}{2}+\frac{\theta(1+\theta)}{2+\theta},
1-\frac{1}{2+\theta})\\
\\
(\theta,1),(1-\frac{1-\theta}{2+\theta},\frac{1}{2+\theta})
\end{array}\right.\right].
\end{eqnarray}
Therefore, the above solution of the fractional equation takes into account 
the Kramers-Moyal result (\ref{kwok}) in the limit $\mu\to 2$ and 
its behaviour in the origin agrees with the master equation. 
The limit $\theta\to 0$ corresponds to the L\'evy process.

\section{Truncated L\'evy flights}

The L\'evy flights are characterized by the power-law tail with the exponent 
smaller than 3 which implies the infinite variance. However, 
in physical systems, which are limited in space, the variance must always 
be finite. The finiteness of the available space must be taken into 
account in any attempt to simulate the random walk in lattices, 
for example, in a model of turbulence which 
describes a transport in a Boltzmann lattice gas \cite{hay}. 
One can constrain the jumping size either by taking into account 
that the Brownian particle actually possesses a finite velocity, 
i.e. to substitute the L\'evy flights by the L\'evy walks, 
or by introducing some cutoff in the jumping size probability distribution. 
As a result, the variance becomes finite and, in the case of the mutually 
independent jumps, the random walk probability distribution converges 
to the Gaussian, according to the central limit theorem. 
The truncation of the L\'evy 
flights can be accomplished either by a simple removing of the tail 
\cite{man} or by multiplying the tail by some fast falling function. 
An obvious choice in this context is the exponential $\exp(-\gamma |x|)$ and 
the algebraic function $|x|^{-\beta}$, where $\beta\ge2-\mu$. The former 
case was considered by Koponen in Ref.\cite{kop}, where an analytic 
expression for the characteristic function was derived. 
The L\'evy flights with exponential truncation serve 
as a model for phenomena in many fields, e.g. in turbulence \cite{dub}, solar 
systems, economy. The distribution
function of velocity and magnetic-field vector differences within 
solar wind can be reasonably fitted in this way \cite{bru}. In the 
framework of the economic research, the L\'evy process is a natural model of 
the financial assets flow and the fractional equations are applied to 
characterize 
the dynamics of stock prices, where rare, non-Gaussian events are frequently 
encountered, in particular if the market exhibits high volatility. 
One can improve in this way the Black-Scholes model, 
commonly used to price the options, which is restricted to the Gaussian 
distributions. In order to incorporate the finiteness of the financial system 
to the fractional equations formalism by means of the exponential truncation 
of the L\'evy tails, models known as CGMY and KoBoL have been devised 
\cite{car}. The non-Markovian fractional equation, which results from the 
CTRW model with the L\'evy distributed and exponentially truncated jump size, 
has been considered in Ref.\cite{car1}. The solutions do not exhibit a typical 
scaling at small time but they converge, asymptotically, to the stretched 
Gaussian which is predicted by the subdiffusive case of CTRW 
with the Gaussian step-size distribution. 
\begin{figure}[tbp]
\includegraphics[width=8.5cm]{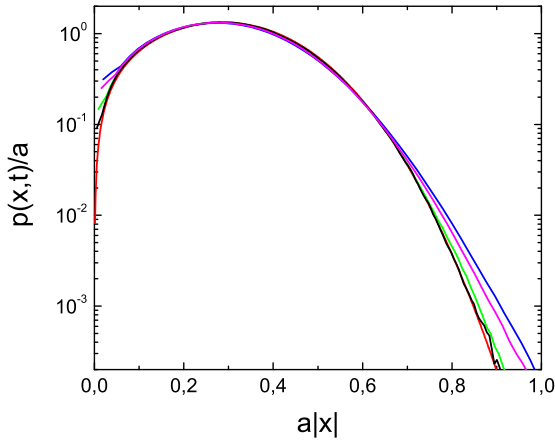}
\caption{The distributions $p(x,t)$ obtained from 
trajectory simulations for the power-law cutoff (\ref{obcp}) with 
$\mu'=5.5$ for $t=5$, $t=10$, $t=50$ and $t=200$ (from top to bottom). 
The distribution predicted by Eq.(\ref{kwok}) is also presented and it 
coincides with the curve for $t=200$. Other parameters are the following: 
$\theta=1$, $\sigma=0.1$ and $a=t^{-1/(2+\theta)}$.}
\end{figure}

In Ref. \cite{kop}, the following 
jumping size distribution, in the form of the L\'evy tail multiplied 
by the exponential factor, has been introduced:
\begin{equation}
\label{obce}
\lambda(x)=N{\mbox e}^{-\gamma |x|}|x|^{-\mu-1} 
\end{equation}
and the characteristic function for the process has been evaluated. 
In the above formula $\gamma\ge 0$ and $N$ is the normalization 
constant. For the symmetric process, the normalized Fourier transform 
from $\lambda(x)$ is given by \cite{kop,car1}
\begin{equation}
\label{obcek}
{\widetilde \lambda(k)}=\frac{4}{\pi}\mu\Gamma(\mu)\tan(\pi\mu/2)[
\gamma^\mu-(k^2+\gamma^2)^{\mu/2}\cos(\mu\arctan(k/\gamma)]+1.
\end{equation}
We keep only the terms of the lowest order in $|k|$. The expansion 
of the expression (\ref{obcek}) produces the following result 
\begin{equation}
\label{obcek1}
{\widetilde \lambda(k)}\approx 1+\frac{2}{\pi}\mu\Gamma(\mu)\gamma^{\mu-2}
\tan(\pi\mu/2)(\mu^2-\mu)k^2\equiv 1-K_E^2 k^2.
\end{equation}
The diffusion process is then described by Eq.(\ref{fraceg}) with $K^2=K_E^2$. 
\begin{figure}[tbp]
\includegraphics[width=8.5cm]{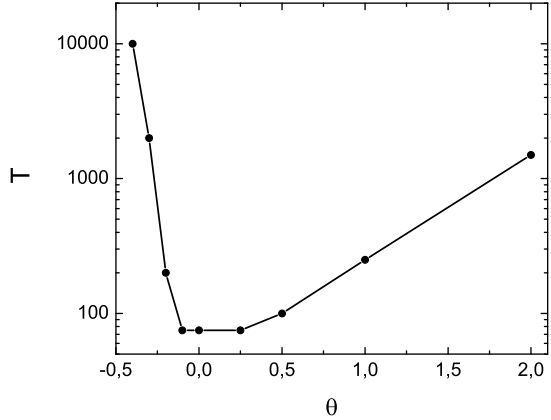}
\caption{The time $T$ needed to reach the distribution (\ref{kwok}), 
as a function of $\theta$, for the case of the algebraic cutoff (\ref{obcp}) 
with the parameters $\mu'=5.5$ and $\sigma=0.1$.}
\end{figure}

On the other hand, one can apply the power-law cutoff to the L\'evy 
tail. In Ref.\cite{sok} the fractional equation of the distributed 
order, with the constant diffusion coefficient, was introduced; 
it implies the cutoff in the form of the power-law function with the 
exponent $5-\mu$. Since the equation involves both the fractional 
and the diffusion component, there is no simple scaling. In this paper, 
we assume the jumping size distribution in the form of the modified 
L\'evy tail: 
\begin{equation}
\label{obcp}
\lambda(x)=\left\{
\begin{array}{ll}
0                             &\mbox{for $|x|\le\sigma$}\\
\mu'\sigma^{\mu'}|x|^{-\mu'-1}&\mbox{for $|x|>\sigma$},
\end{array}
\right.
\end{equation}
where $\mu'=\mu+\beta>2$ to ensure the existence of the second moment. 
The Fourier transform is given by \cite{mar}
\begin{equation}
\label{obcpk}
{\widetilde \lambda(k)}=1-\frac{\mu'}{2(\mu'-2)}(k\sigma)^2-
\left[\Gamma(1-\mu')\cos(\pi\mu'/2)\right](|k|\sigma)^{\mu'}+\dots.
\end{equation}
In this case we have $K^2=\mu'\sigma^2/2(\mu'-2)$, provided we keep 
only the quadratic term. 
\begin{figure}[tbp]
\includegraphics[width=8.5cm]{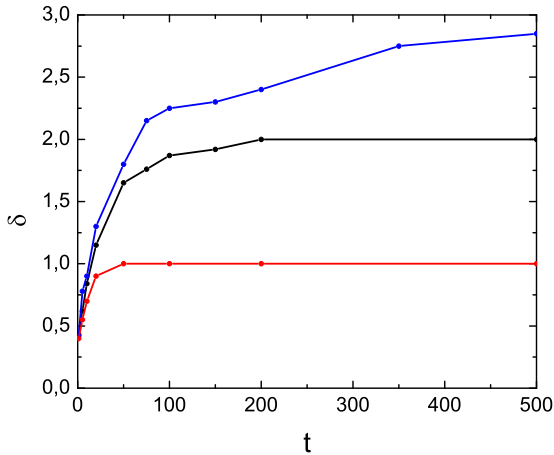}
\caption{The shape parameter $\delta$ of the distribution tail 
$\exp(-const |ax|^\delta)$ as a function of time for the exponential 
cutoff (\ref{obce}). The curves correspond to the cases: $\theta=1$, 
$\theta=0$ and $\theta=-0.5$ (from top to the bottom). The other 
parameters: $\mu=1.5$ and $\gamma=1$.}
\end{figure}

In the following, we evaluate the probability density distribution $p(x,t)$ 
for both forms of the L\'evy flight cutoff, exponential and algebraic, 
by means of the random walk trajectory simulations. The waiting time is sampled 
from Eq.(\ref{poi}) and the jumping size from either (\ref{obce}) or 
(\ref{obcp}). 
Fig.1 presents the power-law case for $\theta=1$, which corresponds to the 
subdiffusion. The results are compared with the Kramers-Moyal limiting 
distribution (\ref{kwok}), which is expected to be reached at large time. 
We observe a rapid convergence for small and medium $|x|$-values, whereas the 
tails reach the form (\ref{kwok}) at about $t=200$. Nevertheless, the shape 
of the tails is always stretched exponential $\sim\exp(-const|ax|^\delta)$ 
and the index $\delta$ rises with time. The speed of convergence to the 
distribution (\ref{kwok}) strongly depends on the parameter $\theta$. In Fig.2 
we present the convergence time $T$ as a function of $\theta$. It is relatively 
short only for small $|\theta|$; for this case the kernel in the master equation 
(\ref{fpkp}) changes weakly with $|x|$ and the higher terms 
in the Kramers-Moyal expansion soon become negligible. $T$ rises rapidly for 
the negative $\theta$: the estimation presented in the figure suggests 
that the dependence $T(\theta)$ is exponential, $\sim\exp(-16\theta)$, which 
yields $T\sim 10^8$ when $\theta$ approaches -1. 
The rapid growth of the time needed to reach convergence of the tails for 
the negative $\theta$ may also be related to a specific shape of the 
distribution. The tails become flat for $\theta<0$, and the asymptotics 
emerges first for very large $|x|$. On the other hand, the probability 
density to stay in the origin, $p(0,t)$, is then infinite: 
we have $p(x,t)\sim t^{-(1+\theta)/(2+\theta)}|x|^\theta$ 
$(|x|\ll1)$, according to Eq.(\ref{kwok}). Note that in the case 
$\theta=0$ we obtain the usual result for diffusion: $p(0,t)\sim1/\sqrt{t}$. 
\begin{figure}[tbp]
\includegraphics[width=8.5cm]{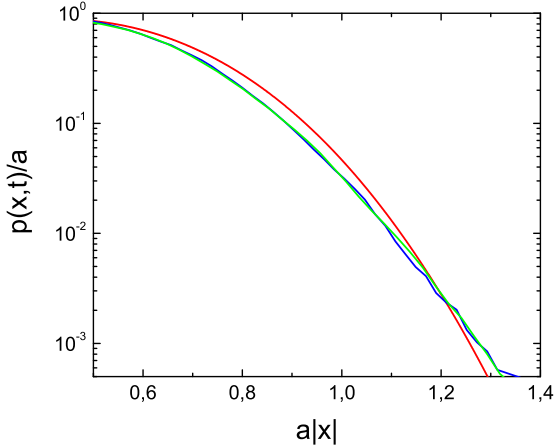}
\caption{The distributions $p(x,t)$ at $t=100$ and $t=200$ obtained from 
trajectory simulations for the exponential cutoff with 
$\mu=1.5$ and $\gamma=2$. These curves coincide and assume the numerically 
estimated shape $\sim a|x|\exp(-4.7(a|x|)^{2.6}$. The distribution predicted by 
Eq.(\ref{kwok}) is also presented. The parameter $\theta=1$ and 
$a=t^{-1/(2+\theta)}$.}
\end{figure}

Application of the exponential cutoff to the L\'evy tail produces similar 
probability density distributions to those presented in Fig.1 and they are also 
characterized by the stretched Gaussian tails. However, the 
convergence rate of the index $\delta$ to the value $2+\theta$, predicted 
by the solution (\ref{kwok}), is smaller than for the power-law truncation 
because 
the kernel in Eq.(\ref{fpkp}) is steeper in this case. 
In Fig.3 we present the dependence $\delta(t)$ for all kinds of the diffusion. 
Initially, the exponent rises fast but then it begins to stabilize and the 
curves 
approach the asymptotic values very slowly, especially for the superdiffusive 
case 
of the negative $\theta$. Discrepancies from Eq.(\ref{kwok}) are more pronounced 
for $\gamma=2$. This case is presented in Fig.4: though the rescaled 
distribution 
seems to be stabilized already for $t=100$, its shape differs 
from (\ref{kwok}) also for intermediate values of $a|x|$. 
\begin{figure}[tbp]
\includegraphics[width=8.5cm]{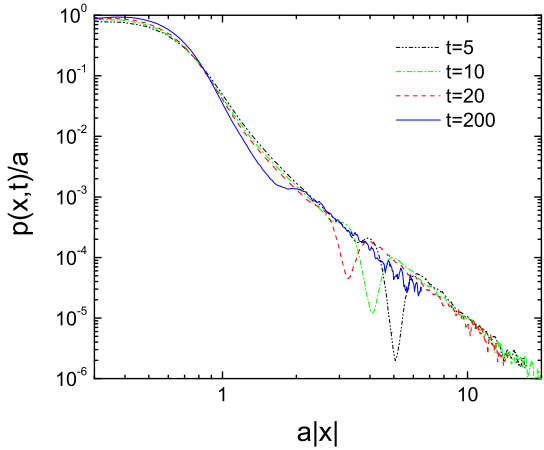}
\caption{The distributions $p(x,t)$, plotted in the double logarithmic scale, 
obtained from trajectory simulations for the power-law cutoff with 
$\mu'=2.5$, $\theta=1$ and $\sigma=0.1$.}
\end{figure}

The exponential asymptotics of the probability density distributions is 
guarantied by the existence of the finite second moment. However, the time 
needed to reach that asymptotics can be so large \cite{man} 
that it cannot be observed 
in any practical realizations of the process. In fact, the convergence rate to 
the normal distribution is governed by the third moment, according to the 
Berry-Ess\'een theorem \cite{fel} which refers to the sum of mutually 
independent 
variables, sampled from the same distribution. It states that if the third 
moment $\rho$ is finite, then the deviation of a given distribution from 
the normal one is less than $33\rho/(4\sigma^3\sqrt{n})$, where $n$ the number 
of steps and $\sigma$ is the standard deviation. If $\rho=\infty$, the 
convergence to the Gaussian may be problematic in practice. To demonstrate 
that the case of divergent third moment is exceptional also for our process, 
let us consider the power-law cutoff of the L\'evy tail, Eq.(\ref{obcp}), 
with the parameter $\mu'=2.5$. The tail of the resulting random walk 
distribution, presented in Fig.5, is no longer exponential but 
it assumes the power-law form $|x|^{-\delta}$ which persists to the largest 
times, 
numerically accessible. The parameter $\delta$ is constant and well determined 
for 
small time: $\delta=3.4$. Moreover, it seems to be independent of $\theta$. The 
presence of algebraic tail is not restricted to the case $\mu'\le3$, for which 
the third moment is divergent; also for slightly larger values of $\mu'$ it is 
clearly visible. In the case $\mu'=3.1$ we find $\delta=4.0$ at small times, 
whereas $\delta=4.5$ for $\mu'=3.5$. 

The probability distributions which possess algebraic tails with $3<\delta<4$ 
are 
of interest in the economic research. It has been suggested 
that such power-laws in financial data arise when the trading behaviour is 
performed in an optimal way \cite{gab}. The stock market data seems to confirm 
that expectation. Extensive studies of the US indexes indicate the power-law 
form of the probability distribution of stock price changes with 
$2.5<\delta<4$ \cite{gop,ple,sta}. Moreover, the distributed order equation of 
Sokolov et al. \cite{sok} predicts the similar value: $\delta=3.3$.

\section{Conclusions}

We have solved the fractional equation which follows from the master equation 
for the jumping process in the diffusion approximation of small wave numbers. 
The jumping size distribution has the L\'evy form. 
The jumping rate depends on position and then the diffusion coefficient 
in the fractional equation is variable. We have considered the jumping rate 
in the algebraic form $\nu(x)=|x|^{-\theta}$, well suited for the 
diffusion on self-similar structures. The generalization to other 
dependences $\nu(x)$ is possible \cite{sro1}. 
It has been demonstrated that the equation is satisfied by the scaling formula 
which can be expressed in terms of the Fox function $H^{1,1}_{2,2}$, when 
one neglects higher terms in the $k$-expansion. 
In the limit $\mu\to2$, the solution reduces 
itself to the Fox function of the lower order and it exhibits the 
stretched-Gaussian asymptotic form (\ref{solex2}). 
The exponent of that function, $1/\alpha$, is related to 
the higher moments and it cannot be uniquely determined 
in the diffusion approximation. The solution predicts all kinds of diffusion, 
both normal and anomalous, which are distinguished by the parameter $\theta$. 
The diffusion equation can 
be solved exactly for the case $\mu=2$ and that solution constitutes the 
Kramers-Moyal approximation of the master equation. The requirement that 
the fractional equation solution should agree with that result in the 
limits $\mu\to2$ and $t\to\infty$ allows us to find additional conditions 
for the Fox function coefficients. 

In the approximation of small wave numbers, 
the problem of truncated L\'evy flights coincides 
with that of the Gaussian jump sizes. We have applied two forms of the 
cutoff: the exponential and the algebraic ones to study the random walk 
process by the Monte Carlo method. In most cases, 
the probability density distributions converge with time to the Kramers-Moyal 
result which predicts $\alpha=1/(2+\theta)$. However, that convergence 
appears very slow if $\theta$ is far away from 0, especially for $\theta<0$. 
If the truncation function is steep, the distribution seems to assume the 
stabilized asymptotic shape which differs slightly from Eq.(\ref{kwok}) 
(see Fig.4) and this conclusion may indicate 
a limit of applicability of the Kramers-Moyal 
approximation. In other cases, form (\ref{kwok}) is actually reached 
after a long time. For smaller time, the scaled distribution is 
time-dependent, in disagreement with the ansatz (\ref{s1}). 
However, the distribution tail is 
always power-law and the exponent $\delta$ depends on time very weakly 
(Fig.3), compared to the function $a(t)$. One can expect that in an experimental 
situation, when the observation time is finite, the distributions are unable 
to converge to (\ref{kwok}) and they may reveal the values of $\delta$ 
smaller than $2+\theta$. 

The convergence of the distribution to (\ref{kwok}) becomes problematic 
not only for very sharp cutoffs but, conversely, for the functions which 
possess a large, in particular infinite, third moment. Numerical simulations 
predict in this case the power-law asymptotics and no trace of 
the exponential tail could be found. The value of the exponent of 
power-law tail agrees with observations, e.g. for the financial data.

\end{document}